# Deduction of the law of motion of the charges from Maxwell equations

A. LOINGER

Dipartimento di Fisica, Università di Milano

Via Celoria, 16, 20133 Milano, Italy

ABSTRACT. — By exploiting suitably a fundamental theorem by Hilbert, we show that the equation of motion of the electric charges is a consequence of Maxwell field equations.

**1.** – As it is well known, in general relativity the law of motion of the electric charges can be deduced from the Einstein-Maxwell field equations (cf. e.g. Dirac, 1975, p.57). On the other hand, according to a *locus classicus*, in the ordinary formulation of Maxwell-Lorentz electrodynamics in Minkowski space-time the law of motion of the charges is independent of the field equations, only the sign of the force density being fixed by the above equations (see e.g. Gupta, 1977, p.6).

We intend to prove that this *locus classicus* is wrong: also in **special** relativity the law of motion of the charges is a consequence of the field equations.

**2.** – The key of our argument is very simple. In general relativity the essential role in the proof in question is played by Bianchi identities for the Ricci-Einstein tensor. Now, by virtue of a luminous theorem by Hilbert (1915), these identities can be viewed as a consequence of the invariance of the action integral under a change of co-ordinates that leaves the bounding surface unchanged (see e.g. Schrödinger, 1960, p.93, and Dirac, 1975, p.59). More generally, an invariance property of this kind assures the validity of the differential conservation law for any energy-momentum tensor. In a recent paper (Loinger, 1997) we have observed that Hilbert theorem can be also applied to the formulation of *special* relativity in *general curvilinear* co-ordinates. We shall now exploit this remark to demonstrate our thesis.

**3.** − Let us consider a charged "dust" of invariant mass density ρ and invariant charge density σ. Adopting Dirac's notations (Dirac, 1975), the energy-momentum tensor of the insulated system "dust" plus the electromagnetic field generated by the charges in motion is given by

$$(3.1) \qquad W^{\mu\nu} = \rho \, v^\mu v^\nu + E^{\mu\nu} \; ,$$

where $v^\mu = dx^\mu/ds$, and $ds^2 = g_{\mu\nu} \, dx^\mu dx^\nu$ ; $g_{\mu\nu} \, (x^0, x^1, x^2, x^3)$ is the metric tensor of Minkowski space-time referred to a generic frame of co-ordinates; $E^{\mu\nu}$ is the energy-momentum tensor of the e.m. field $F^{\mu\nu}$:

$$(3.2) \qquad 4\pi \, E^{\mu\nu} = - F^\mu_{\;\alpha} F^{\nu\alpha} + \tfrac{1}{4} g^{\mu\nu} F_{\alpha\beta} F^{\alpha\beta} \; .$$

(Remark that $\rho \, v^\mu v^\nu$ is the energy-momentum tensor of an uncharged "dust"; its conservation law yields the obvious equation of motion of the "dust" particles.)

The Maxwell field equations for our continuum are:

$$(3.3) \qquad F_{\mu\nu:\sigma} + F_{\nu\sigma:\mu} + F_{\sigma\mu:\nu} = F_{\mu\nu,\sigma} + F_{\nu\sigma,\mu} + F_{\sigma\mu,\nu} = 0 \; ;$$

$$(3.4) \qquad F^{\mu\nu}_{\;\;\;:\nu} = 4\pi \, \sigma \, v^\mu ,$$

where the colon represents the covariant derivative and the comma the ordinary partial derivative.

Now, the above extension of Hilbert theorem (Loinger, 1997), assures us that the conservation law

$$(3.5) \qquad W^{\mu\nu}{}_{:\nu} = 0$$

is *fully* independent of Maxwell equations and of the equation of motion of the charges. But if we take into account eqs. (3.3) and (3.4), we obtain from eq. (3.5) with an easy calculation – in a complete formal analogy with the procedure of general relativity – the dynamical equation

$$(3.6) \qquad \rho \, v_{\mu \,:\nu} \, v^{\nu} + \sigma \, F_{\mu\nu} \, v^{\nu} = 0 \,,$$

i.e. the law of motion of the charges. *Q.e.d.*

(From an *anschaulich* point of view this result is not too astonishing: indeed, we could start from the Einstein-Maxwell field equations of general relativity written for an *evanescent* constant of universal gravitation .....).

*————————*